\documentclass[journal=jctcce, articletitle=true]{achemso}
\setkeys{acs}{articletitle = true}
\SectionNumbersOn

\usepackage[english]{babel}
\usepackage{graphicx}
\usepackage{amssymb}
\usepackage{amsmath}
\usepackage[colorlinks,citecolor=blue,urlcolor=blue]{hyperref}
\usepackage{mathtools}
\usepackage{amsfonts}
\usepackage{color}
\usepackage[update,prepend]{epstopdf}
\usepackage{pifont}
\newcommand{\cmark}{\ding{51}}%
\newcommand{\xmark}{\ding{55}}
\usepackage[toc,page]{appendix}

\usepackage{colortbl,booktabs}%
\newcommand{\myrowcolour}{\rowcolor[gray]{0.925}}
\newcolumntype{?}{!{\vrule width 0.4pt}}

\newcommand{\la}{\left\langle}
\newcommand{\ra}{\right\rangle}
\def \figwidth {10cm}

\title{BoltzmaNN: Predicting effective pair potentials and equations of state using neural networks}

\author{Fabian Berressem}
\affiliation{Institute of Physics, Johannes Gutenberg University Mainz, Staudingerweg 7, 55128 Mainz, Germany}

\author{Arash Nikoubashman}
\email{anikouba@uni-mainz.de}
\affiliation{Institute of Physics, Johannes Gutenberg University Mainz, Staudingerweg 7, 55128 Mainz, Germany}

\date{\today}

\begin{document}

\begin{abstract}
Neural networks (NNs) are employed to predict equations of state from a given isotropic pair potential using the virial expansion of the pressure. The NNs are trained with data from molecular dynamics simulations of monoatomic gases and liquids, sampled in the $NVT$ ensemble at various densities. We find that the NNs provide much more accurate results compared to the analytic low-density limit estimate of the second virial coefficient. Further, we design and train NNs for computing (effective) pair potentials from radial pair distribution functions, $g(r)$, a task which is often performed for inverse design and coarse-graining. Providing the NNs with additional information on the forces greatly improves the accuracy of the predictions, since more correlations are taken into account; the predicted potentials become smoother, are significantly closer to the target potentials, and are more transferable as a result. 
\end{abstract}

\maketitle

\section{Introduction}
Understanding and predicting the relationship between the (macroscopic) properties of a material and its (microscopic) building blocks is one of the key challenges in materials research and physics. One important goal in statistical physics is the accurate prediction of the phase behavior on the basis of the (effective) pair potential $U(r)$ acting between the particles. According to van der Waals' theorem of corresponding states, all simple fluids obey the same reduced equation of state (EOS), if the thermodynamic variables are rescaled by their value at the critical point. However, this law only applies for systems with conformal pair potentials, {\it i.e.}, when the potentials can be fully superimposed by adjusting the interaction strength and particle diameter which is rarely the case in practice. Noro and Frenkel extended this principle by including the reduced second virial coefficient for quantifying the effective range of the attraction.\cite{noro:jcp:2000} This extended approach can provide accurate predictions for pair potentials which are characterized by attractive interactions with ranges much smaller than the particle size,\cite{noro:jcp:2000, lu:nature:2008} but it is expected to fail for more complex pair potentials which, {\it e.g.}, include a repulsive barrier. Thus, alternative prediction tools are highly desirable, especially given that a large number of (effective) pair potentials in soft matter are bounded or have repulsive barriers.\cite{bolhuis:jcp:2001, likos:physrep:2001, sciortino:prl:2004, mossa:langmuir:2004, mladek:prl:2008}

Progress in this field has wide implications, not just in terms of our fundamental understanding, but also due to the large number of potential technological applications. Various mechanical, optical, and electronic material properties critically depend on the degree of ordering of their atomic or (macro)molecular constituents. In the conventional {\it forward design} approach, the development of new materials typically begins with designing candidate building blocks that are expected to lead to the desired properties. Then these candidates are created, tested, and, if necessary, modified, until the compound with the desired properties has been identified. However, this iterative procedure can quickly become rather time- and resource consuming. Therefore, good initial candidates are required to achieve convergence in a reasonable time frame. Due to these inherent issues, there has been a recent paradigm shift towards the {\it inverse design} process, where the building blocks are inferred from the desired target properties. This pathway has been explored for a range of soft materials,\cite{jain:aiche:2014, sherman:jcp:2020} including athermal granular media,\cite{jaeger:sm:2015} colloids,\cite{lindquist:jcp:2016, chen:jpcb:2018} and block copolymers.\cite{khadilkar:mm:2017, gadelrab:msde:2017} In recent years, informatics-driven approaches have gained popularity that utilize machine learning algorithms on large databases to identify previously unrecognized patterns and to predict new candidate materials.\cite{behler:jcp:2016, audus:ml:2017, ferguson:jpcm:2017, bereau:jcp:2018, bereau:book:2018, jackson:coce:2019, Wu2019, Schmidt2019}

This inverse design process is strongly related to the problem of top-down coarse-graining, where the goal is to derive effective (pair) potentials for a system with a reduced number of degrees of freedom, while preserving selected target properties of the original fine-grained representation. Several methods have been developed for this task, including Reverse Monte Carlo (RMC),\cite{RMC, keen:nature:1990, lyubartsev:pre:1995} Iterative Boltzmann Inversion (IBI),\cite{reith:jcc:2003} and simulated annealing-based optimization.\cite{rechtsman:prl:2005} These techniques have been employed successfully for, {\it e.g.}, developing effective pairwise potentials from experimental structure measurements\cite{reatto:pra:1986, mueller:langmuir:2014, song:jpcl:2017} and for coarse-graining atomistic simulations.\cite{mp:cpc:2002, reith:jcc:2003, milano:jpcb:2005, bayramoglu:mm:2012} Achieving transferability and representability of such coarse-grained models is, however, a key challenge,\cite{louis:jpcm:2002, rosenberger:epj:2016} given that the multi-body potential of mean force is usually approximated by an effective pair potential.\cite{RMC, keen:nature:1990, lyubartsev:pre:1995, louis:jpcm:2002, reith:jcc:2003, rechtsman:prl:2005, shell:jcp:2008, rosenberger:epj:2016} Further, although the IBI scheme should in principle provide a unique solution for a given $g(r)$,\cite{henderson:pla:1974} in practice however, convergence of this iterative procedure is not guaranteed.\cite{rosenberger:epj:2016} Various strategies have been devised for improving these and other aspects of coarse-graining, such as the addition of thermodynamic constraints,\cite{reith:jcc:2003, wang:epje:2009, das:jcp:2010} and the development of improved methods is an active field of research.\cite{Bereau2018, Lebold2019, Scherer2020, Meinel2020, Foley2020}

In this work, we employ artificial neural networks (NNs) for predicting the EOS from a given isotropic pair potential $U(r)$ using the virial expansion of the pressure. Further, we design and train NNs for computing (effective) pair potentials from a given radial pair distribution function, $g(r)$. The training and test data for the NNs are generated from Molecular Dynamics (MD) simulations of monoatomic gases and liquids in the canonical ensemble. The NNs developed in this work as well as example scripts are available online.\cite{boltzmann} The rest of this manuscript is organized as follows. In Sec.~\ref{sec:methods}, we provide a brief summary of the numerical methods we used as well as how the data were generated. In Sec. ~\ref{sec:results}, we present our results, where Sec.~\ref{sec:b2effective} focuses on using NNs for determining the EOS from a given pair potential, and Sec.~\ref{sec:potpredictor} discusses how NNs can be used for computing (effective) pair potentials from the radial pair distribution function. Section ~\ref{sec:conclusions} contains our conclusions and a brief outlook

\section{Methodology}
\label{sec:methods}
\subsection{Design of Neural Networks}
\label{subsec:NN}
We used three different NN architectures in this work, {\it i.e.}, fully connected dense NNs (DN), convolutional NNs (CN), and U-Nets\cite{UNet} (UN). The DNs consist of layers of neurons, where all neurons between two subsequent layers are connected to each other. Operations in DNs are limited to simple matrix-matrix and matrix-vector-multiplications, combined with (nonlinear) activation functions to break linearity. The CNs use convolutional operations which, compared to a DN, drastically reduce the number of parameters necessary by considering only local correlations between points. Therefore, CNs are useful for strongly correlated data, as is typically the case in image processing but also in our problem. In UNs, the input is first processed using convolutional layers then upsampled again and concatenated to a former stage of the processing, hence extracting features and combining them with the original input for further processing.

The accuracy of an NN depends strongly on the information that is provided to it, {\it e.g.}, the average particle number density and/or the force. Also, the format in which this information is represented plays an important role. For instance, one could use $\exp\left[-\beta U(r)\right]$ rather than $U(r)$, which may be favorable for pair potentials with a strong repulsion combined with an attraction of significantly smaller strength. A detailed study on the effect of these parameters is provided in Sec.~\ref{sec:results}.

Another key aspect of the NNs is the loss function, which essentially controls the properties that should be optimized by the NN. The loss functions used in this work were combinations of three basic loss functions, the squared error (SE), the absolute error (AE), and the logcosh error (LE). For a given pair of target and prediction, $(y, \hat{y})$, the basic losses are
\begin{align}
	l_{\rm SE}(y, \hat{y}) &= \left(y - \hat{y}\right)^2 ,\\
	l_{\rm AE}(y, \hat{y}) &= \left|y - \hat{y}\right| ,\\
	l_{\rm LE}(y, \hat{y}) &= \ln\left[\cosh \left(y - \hat{y}\right)\right] .
\end{align}
The loss of the entire output vector is then determined as the mean over the losses of the individual nodes. For instance, the mean squared error (MSE) is given by
\begin{align}
	L_{\rm MSE}(\mathbf{y}, \mathbf{\hat{y}}) &= \frac{1}{d} \sum_{i=1}^{d} l_{\rm SE}(y_i, \hat{y}_i), 
	\label{eq:LMSE}
\end{align}
with $\mathbf{y}, \mathbf{\hat{y}} \in \mathbb{R}^d$ and dimensionality $d$ of the output vector ({\it e.g.} $d=50$ for predicting pair potentials). Analogous expressions are used for $L_{\rm MAE}$ and $L_{\rm MLE}$. We indicate the average loss taken over multiple output vectors through angular brackets, $\la \dots \ra$.

Expressions like Eq.~(\ref{eq:LMSE}) measure the difference between two single points of $\mathbf{y}$ and $\mathbf{\hat{y}}$, but do not capture the correlations between neighboring points. To ensure that the predicted potentials vary smoothly with $r$, it is therefore helpful to include a Laplace-like term as well as loss terms correlating the differences of the potential at different distances
\begin{equation}
	L_U =  L_{\rm MLE} + \alpha_\Delta L_\Delta + \sum_{k=1}^{4} L_k \alpha^k 
	\label{eq:customloss}
\end{equation}
with weight $\alpha_\Delta$ chosen as $\alpha_\Delta = 2$. The terms $L_k$ can optionally be multiplied by a factor $\alpha \leq 1$ to reduce their contribution with increasing distance ($\alpha = 1$ was used in this work). The term $L_\Delta$ is the discretized Laplace term, which in the case of $l_{\rm SE}$ reads
\begin{equation}
	L_\Delta = \frac{1}{d-2} \sum_{i=1}^{d-2} l_{\rm SE}\left(y_{i+2} - 2 y_{i+1} + y_i, \hat{y}_{i+2} - 2\hat{y}_{i+1} + \hat{y}_i\right).
\end{equation}

The loss terms $L_k$ in Eq.~(\ref{eq:customloss}) are given by the loss function applied to the difference of potential values being $k$ indices apart. For example, with $l_{\rm SE}$ this term reads
\begin{equation}
	L_k = \frac{1}{d-k} \sum_{i=1}^{d-k} l_{\rm SE}\left(y_{i+k} - y_i, \hat{y}_{i+k} - \hat{y}_i \right) .
	\label{eq:distance-msd}
\end{equation}
To see how this definition naturally introduces spatial correlations, one can rewrite $L_k$ from Eq.~(\ref{eq:distance-msd}) as follows
\begin{align}
	\label{eq:distance-msd2}
	L_k =& \frac{1}{d-k} \sum \limits_{i=1}^{d-k} \big[  l_{\rm SE}\left(y_{i+k}, \hat{y}_{i+k}\right) + l_{\rm SE}\left(y_{i}, \hat{y}_{i}\right) \\
	\nonumber
	& - 2\left(y_{i+k} - \hat{y}_{i+k}\right) \left(y_{i} -  \hat{y}_{i}\right)\big] .
\end{align}
In this representation of $L_k$, it is clear that the last term in Eq.~(\ref{eq:distance-msd2}) introduces correlations between values at different distances.

All NNs have been constructed and trained using Tensorflow v. 1.13.1.\cite{tensorflow} The networks were trained using an Adam (adaptive moment estimation) optimizer\cite{ADAM} with a learning rate of 0.001. In order to find a suitable NN topology, an extensive grid search was conducted for the three network architectures, where we systematically studied the influence of the loss function as well as the number and shape of the layers, {\it i.e.}, the width or number of filters and kernel size. During this grid search, the NNs were trained for $10^4$ epochs without any optional parameters (see Sec.~\ref{sec:results} below). For each architecture, we chose the NN with the lowest $\la L_{\rm MSE} \ra$ and $\la L_{\rm MAE} \ra$ evaluated through 4-fold cross-validation, and trained these NNs then for $2 \times 10^5$ epochs with optional parameters. For the final benchmarks, the selected NNs were trained for $5 \times 10^5$ epochs using all data from the training and validation sets.

\subsection{Generation of data}
\label{subsec:data}
In order to generate training and test data for the NNs, we performed a series of MD simulations of monoatomic gases and liquids in the canonical ensemble, and determined the resulting pressures $P$ and radial pair distribution functions $g(r)$. If not stated otherwise explicitly, the particle diameter $\sigma$ and the thermal energy $k_{\rm B} T$ are used as the units of length and energy, respectively. Pair potentials $U(r)$ were generated as spline functions with cutoff distances $r_{\rm cut} \in (0, 5\,\sigma]$. The number of base points was chosen randomly in the range $n \in [6, 10]$, and the base points were distributed randomly at distances in the range $[0, r_{\rm cut}]$ with magnitudes in the interval $[0, 15\,k_{\rm B}T]$. Then a smoothing spline function was applied to connect these points. The degree of the spline function was randomly drawn between $k = 2$ and $k = 5$ to allow for a large variety of shapes [note that $k \geq 2$ is required to guarantee differentiability of $U(r)$]. To ensure physically meaningful pair potentials, the splines were fitted with the constraint $F(r_{\rm cut}) = -\frac{\partial}{\partial r} U(r)\vert_{r_{\rm cut}} = 0$. Further, the potentials were shifted so that $U(r_{\rm cut}) = 0$. Hard-core interactions were included in selected pair potentials by adding the Weeks-Chandler-Andersen (WCA) potential\cite{weeks:jcp:1971} with diameter $\sigma$ and interaction strength $\varepsilon = k_{\rm B}T$.

\begin{figure}[htbp]
	\includegraphics[width=\figwidth]{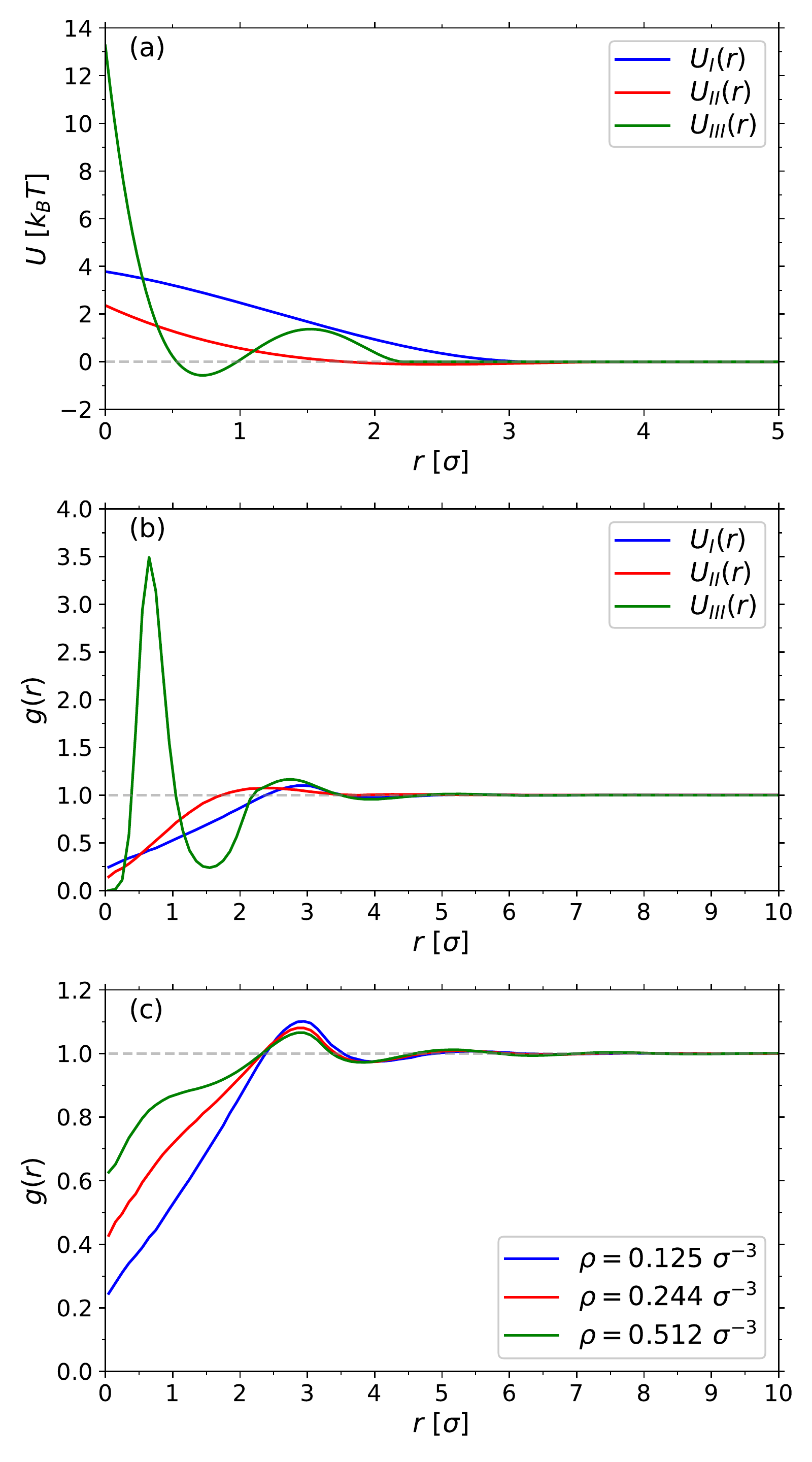}
	\caption{(a) Three selected pair potentials $U(r)$ used in this work, and (b) corresponding	radial pair distribution functions, $g(r)$, recorded at particle number density $\rho = 0.125\,\sigma^{-3}$. (c) $g(r)$ for $U_{\rm I}$ at three investigated values of $\rho$, as indicated.}
	\label{fig:samplepot}
\end{figure}

Using this procedure, 657 potentials were generated, where 344 of them had an additional hard-core contribution. For the MD simulations, tabulated potentials were created with 200 evenly distributed points in the range $[0, r_{\rm cut}]$. For the optimization of the NNs, these tables were then further downsampled to 50 points, evenly distributed in the interval $[0, 5\,\sigma]$. Figure~\ref{fig:samplepot}(a) shows a selection of potentials which were generated using our procedure, while the resulting radial pair distribution functions are plotted in Fig.~\ref{fig:samplepot}(b,c).

The MD simulations were conducted in a cubic simulation box with edge length $L = 40\,\sigma$ and periodic boundary conditions in all directions. The temperature was fixed at $T = 1$ using a Langevin thermostat, and the equations of motions were solved using the standard Velocity-Verlet algorithm with a time step of $\Delta t = 0.005 \tau _{\rm MD}$, with intrinsic MD unit of time $\tau_{\rm MD} = \sqrt{ m \sigma^2 /(k_B T)}$. Simulations were conducted at four different particle number densities, {\it i.e.}, $\rho = 0.125\,\sigma^{-3}$ ($N = 8000$), $\rho = 0.244\,\sigma^{-3}$ ($N = 15625$), $\rho = 0.512\,\sigma^{-3}$ ($N = 32768$), and $\rho = 1.0\,\sigma^{-3}$ ($N = 64000$). Each simulation was first run for $5 \times 10^6$ MD steps for equilibration, and then for additional $5 \times 10^6$ MD steps to sample $P$ and $g(r)$ [see Fig. \ref{fig:samplepot}(b,c) for examples]. For the discretization of the calculated $g(r)$ we chose 200 points that were evenly distributed in the interval $[0, L/2]$. In total, 1678 simulations were conducted, with 529 containing potentials with hard-core repulsion. Simulations were discarded from our analysis, which did not reach equilibrium in the allotted time or formed heterogeneous structures [see ESI for more information], leading to a final number of 891 data points. Of those, 790 data points were used as training and validation sets, while the remaining 101 were reserved for testing only. All simulations were performed on GPUs using the HOOMD-blue software package (v. 2.4.2).\cite{hoomd1, hoomd2}

\section{Results}
\label{sec:results}

\subsection{Predicting the equation of state}
\label{sec:b2effective}
We applied NNs for predicting the pressure $P$ at a given density from the employed pair potential {\it via} three different strategies:
\begin{enumerate}
\item The pair potential is directly mapped to the pressure, $U(r) \mapsto \beta P$;
\item The pair potential is first mapped to an effective, density-dependent second virial coefficient $U(r) \mapsto B_2^*$ and then to the pressure, $\beta P = \rho + \rho^2 B_2^*$;
\item The pair potential is first mapped to a set of $n$ virial coefficients $U(r) \mapsto \mathcal{B}_n \coloneqq \{B_i\; | \; 2 \leq i \leq 5 \}$ and then to the pressure, $\beta P = \rho + \sum_{B_i \in \mathcal{B}_n} B_i \rho^i$.
\end{enumerate}

\begin{figure}[htbp]
	\includegraphics[width=\figwidth]{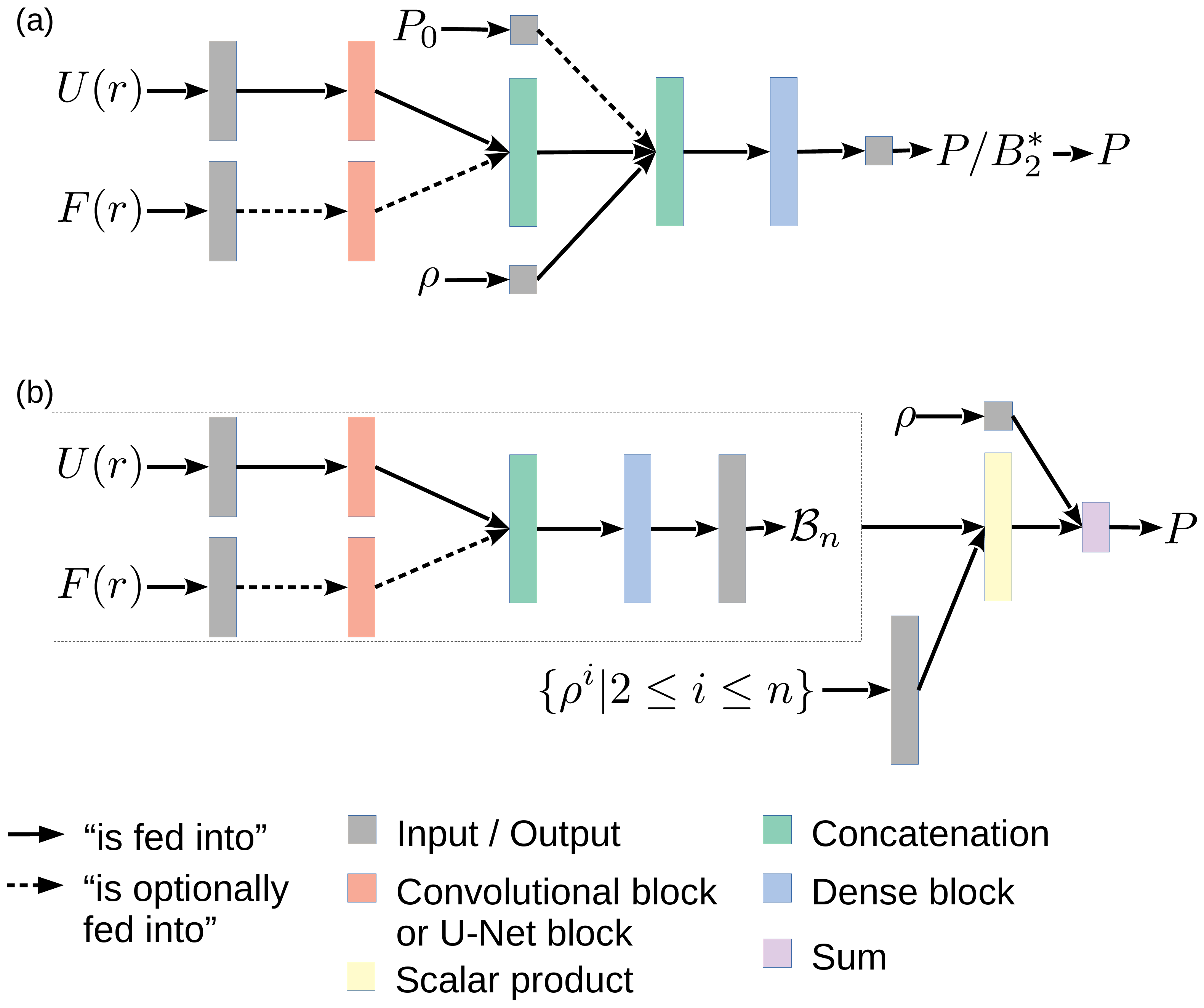}
	\caption{Schematic representation of the NNs used for computing the pressure $P$ (a) directly or {\it via} the effective second virial coefficient, $B_2^*$, and (b) through a set of $n=5$ density-independent virial coefficients $\mathcal{B}_n = \left\{B_i \right\}$. The dotted box in (b) indicates the central part of the NN, while the operations outside of this box do not contain learnable parameters and are only used to calculate $P$.}
	\label{fig:NN_schematic_B2}
\end{figure}

Figure~\ref{fig:NN_schematic_B2} shows schematic representations of the NNs used for predicting $P$ from a given pair potential $U(r)$ and density $\rho$. For the strategies $U(r) \mapsto \beta P$ and $U(r) \mapsto B_2^*$, the density was included as an additional input after convolutional operations, so that the density-dependence is learned by the NN. The calculation of $P$ for the network calculating $\mathcal{B}_n$ is achieved {\it via} layers calculating the scalar product or sum as indicated in the schematic. These layers do not contain any adjustable parameters, so that the learned virial coefficients $\mathcal{B}_n$ are density independent, while still allowing for the loss function to be applied to the density-dependent pressure $P$.

The only required input parameters of our NNs are $U(r)$ and $\rho$. In principle, one could directly use the tabulated pair potentials $U(r)$  as an input of the NNs. In practice, however, this might cause numerical issues for strongly repulsive potentials, since a further increase of the interaction strength beyond a certain threshold has little or no effect. Therefore, it is sensible to reduce the search space and minimize the risk of overfitting by limiting the maximum repulsion of $U(r)$ during the optimization of the NNs. The range of the pair potential $U(r)$ was optionally constrained either by representing it in exponential space, {\it i.e.}, $\exp\left[-\beta U(r)\right]$, or by using a clipped potential, $\bar{U}(r) = \max\left[U(r), U_{\rm cut}\right] \forall r$ with $U_{\rm cut} = 20\,k_{\rm B}T$. The choice of $U_{\rm cut}$ is somewhat arbitrary but our tests indicated that the specific value does not significantly alter the outcome as long as $U_{\rm cut}$ separates soft and hard potentials. 

Furthermore, we tested how additional information can improve the prediction accuracy. The force $F(r)$ was optionally included in the NNs, by applying the same operations as for $U(r)$ and then concatenating the outputs for further processing in fully dense layers (see Fig.~\ref{fig:NN_schematic_B2}). If $\bar{U}(r)$ was used as the input, then the force was calculated before clipping to avoid unphysical discontinuities. If the input was $\exp\left[-\beta U(r)\right]$, then $-\frac{\partial}{\partial r} \exp\left[-\beta U(r)\right]$ was used as the force input. Note that including the derivative almost doubles the number of nodes in the network. For the NNs learning $B_2^*$ or $P$ directly, we optionally provided also a pressure estimate $P_0$, calculated using the analytic second virial coefficient $B_2$. In a homogeneous system with isotropic interactions, $B_2$ is readily available from $U(r)$\cite{hansen:book:2006}
\begin{equation}
	B_2(T) \approx -2\pi \int_0^\infty f(r, T) r^2\ {\rm d}r ,
	\label{eq:B2_isotropic}
\end{equation}
with Mayer $f$-function
\begin{equation}
	f(r, T) = \exp\left[ -\beta U(r) \right] - 1 .
\end{equation}
Thus for $\rho \to 0$, the pressure $P_0$ can be directly computed from the pair potential $U(r)$
\begin{equation}
	\beta P_0 = \rho - 2\pi\rho^2 \int_0^\infty f(r, T) r^2\ {\rm d}r .
	\label{eq:virial_approximation}
\end{equation}

In what follows, we will use the following naming convention for the NNs: For a given architecture X and given options y, we name the network X/y, {\it e.g.},  for a UN with force information we use UN/f. The abbreviations for the different combinations are summarized in Table~\ref{tab::abbreviations_B2}. Hence, we investigated 90 different combinations of network architectures and optional parameters. It is clear that optimizing the hyperparameters (see Sec.~\ref{subsec:NN}) for each individual case is computationally infeasible, and therefore we optimized only the DN/x, CN/x, and UN/x networks. Then, those optimized hyperparameters are used for all other networks of the same architecture to systematically investigate the effect of the additional information provided to the NN as well as the representation of the input and output data. In most cases, the variation of the hyperparameters resulted in changes of the prediction accuracy on the order of the variation between the folds, thus typically less than the effect of the parameters listed in Table~\ref{tab::abbreviations_B2}. Only in selected cases, we found that the NNs were too small to capture the relevant details or too large to cause overfitting.

\renewcommand{\arraystretch}{1.1}
	\begin{table}[htbp]
 	\begin{tabular}{c|c|c|c|c}
 		\toprule
 		Abbreviation & $\bar{U}(r)$ & $\exp[-\beta U(r)]$ & force  & $P_0$ \\ 
 		\cmidrule[0.4pt](r{0.125em}){1-5}
		x & \xmark & \xmark & \xmark & \xmark\\
		\myrowcolour
		c & \cmark & \xmark & \xmark & \xmark\\
		e & \xmark & \cmark & \xmark & \xmark\\
		\myrowcolour		
		f & \xmark & \xmark & \cmark & \xmark\\
		i & \xmark & \xmark & \xmark & \cmark\\
		\myrowcolour		
		cf & \cmark & \xmark & \cmark & \xmark\\
		ci & \cmark & \xmark & \xmark & \cmark\\
		\myrowcolour		
		ef & \xmark & \cmark & \cmark & \xmark\\
		ei & \xmark & \cmark & \xmark & \cmark\\
		\myrowcolour		
		fi & \xmark & \xmark & \cmark & \cmark\\
		cfi & \cmark & \xmark & \cmark & \cmark\\
		\myrowcolour		
		efi & \xmark & \cmark & \cmark & \cmark\\
		\bottomrule
 	\end{tabular} 
 	\caption{List of abbreviations for the optional parameters of the networks.}
 	\label{tab::abbreviations_B2}
 \end{table} 
\renewcommand{\arraystretch}{1.0}

\begin{figure*}[htbp]
	\includegraphics[width=\textwidth]{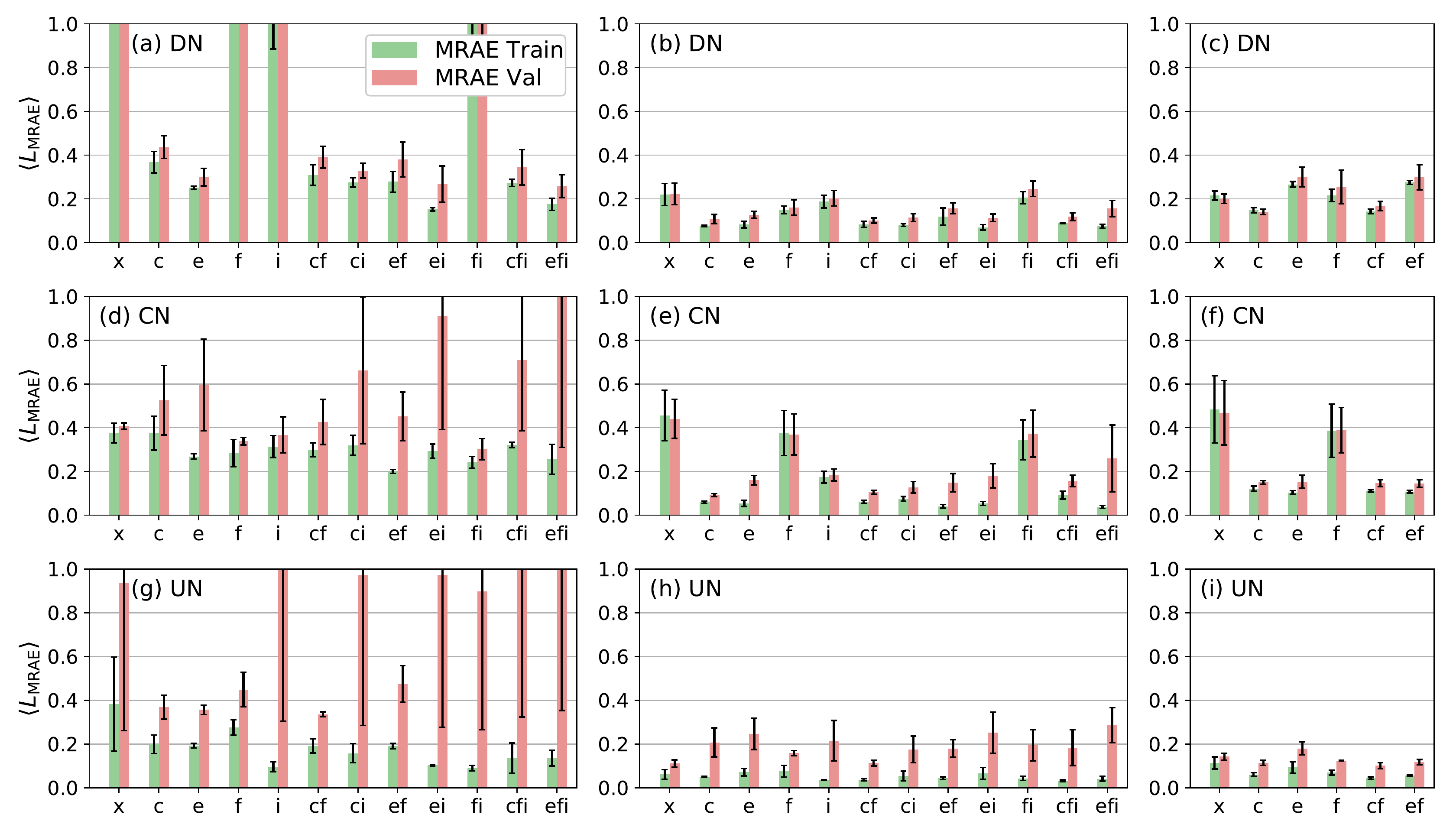}
	\caption{Mean relative absolute error $\la L_{\rm MRAE} \ra$ between predicted pressures $\hat{P}$ and target pressures $P$ for different NN architectures and options. The panels are ordered from top to bottom as DN, CN and UN and from left to right as  mapping to $P$, mapping to $B_2^*$ and mapping to $\mathcal{B}_n$.}
	\label{fig:LossPlotB2}
\end{figure*}

Figure~(\ref{fig:LossPlotB2}) shows the mean relative absolute error between the predicted and target pressures, $\la L_{\rm MRAE} \ra = \la\left|P - \hat{P}\right|/P\ra$, for all investigated NNs. The errorbars in Fig.~\ref{fig:LossPlotB2} were calculated as the standard error of the mean over all four folds. For comparison, the mean relative absolute error between the target pressure $P$ and the pressure $P_0$ estimated {\it via} $B_2$ was $\la L_{\rm MRAE} \ra \approx 1.4$. In most cases, the NNs provide significantly more accurate predictions compared to $P_0$, which is reasonable given that Eq.~(\ref{eq:virial_approximation}) is strictly valid only in the limit $\rho \to 0$. Further, the NNs that directly map $U(r) \mapsto \beta P$ perform worse than the NNs which first predict $B_2^*$ or $\mathcal{B}_n$ and then compute $P$.

Using either the clipped potential $\bar{U}(r)$ or $\exp[-\beta U(r)]$ as the input instead of $U(r)$ typically leads to better results and reduces overfitting in most of the cases. In contrast, including the forces did not improve the performance notably, but was even detrimental in some cases, as can be seen, for example, by comparing the performance of CN/fi and CN/i predicting $B_2^*$ [see Fig.~\ref{fig:LossPlotB2}(e)]. We surmise that this behavior occurs because including the force does not provide any crucial additional information compared to the pair potential, while increasing the complexity of the network architecture and thus increasing the risk of overfitting. Providing $P_0$ sometimes improved performance, but the effect was not as pronounced compared to the other parameters. For the UNs predicting $B_2^*$ [Fig.~\ref{fig:LossPlotB2}(h)], including $P_0$ resulted in a consistently lower accuracy, which was likely due to overfitting as indicated by the small loss on the training set compared to the loss in the validation set throughout.

Based on the comparison shown in Fig.~\ref{fig:LossPlotB2}, we chose UN/c predicting $\mathcal{B}_n$ as the final network because of the following reasons: The pressures  predicted by this NN were among the most accurate with small mean relative absolute errors $\la L_{\rm MRAE} \ra \approx 0.06 \pm 0.01$ and $\approx 0.11 \pm 0.01$ for the training and test set, respectively. Further, the mapping $U(r) \mapsto \mathcal{B}_n$ should also generalize better to densities which were not part of the training set, as the predicted $\mathcal{B}_n$ are explicitly density independent, with the only density dependence of $\hat{P}$ coming from the virial expansion. Further, we used clipped potentials $\bar{U}(r)$, which guaranteed that the values provided to the NN are bounded, leading to better generalizability and less overfitting. Forces were not included in the final network, because they did not always improve the prediction accuracy, but increased the model complexity and the risk of overfitting.

\begin{figure}[htbp]
	\includegraphics[width=\figwidth]{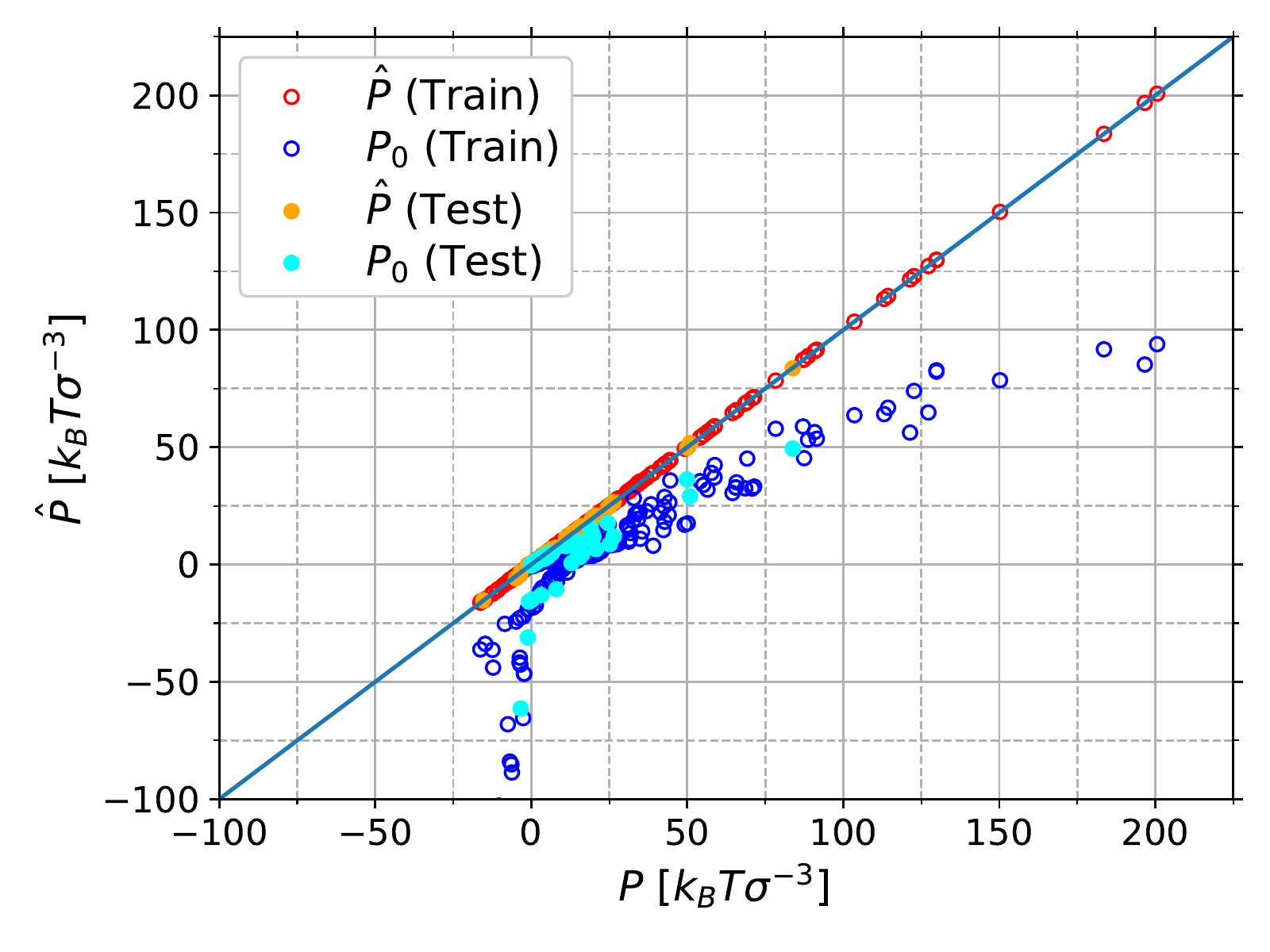}
	\caption{Predicted pressure $\hat{P}$ from UN/c (red and orange symbols) and $P_0$ estimated from  Eq.~(\ref{eq:virial_approximation}) (blue and cyan symbols) {\it vs.} the target pressure $P$. Open and filled symbols show results from the training and test set, respectively.}
	\label{fig:PerformanceB2}
\end{figure}

For the final benchmark, all data (except those in the test set) were used for training the NN. Figure~\ref{fig:PerformanceB2} shows the pressure $\hat{P}$ predicted by the UN/c network {\it vs.} the target pressure $P$. Here, we have also included the low-density estimate $P_0$ [see Eq.~(\ref{eq:virial_approximation})]. The predicted pressures $\hat{P}$ are very close to the target values $P$, with coefficients of determination $R^2 \approx 1.00$ for both the training and test set. In contrast, $P_0$ deviated strongly from $P$, especially for the systems at high density and/or with $P<0$. This rather poor agreement with the target pressure is reflected in the coefficients of determination, which was $R^2 \approx 0.172$.

\subsection{Potential prediction from pair distribution function}
\label{sec:potpredictor}
\begin{figure}[htbp]
	\includegraphics[width=\figwidth]{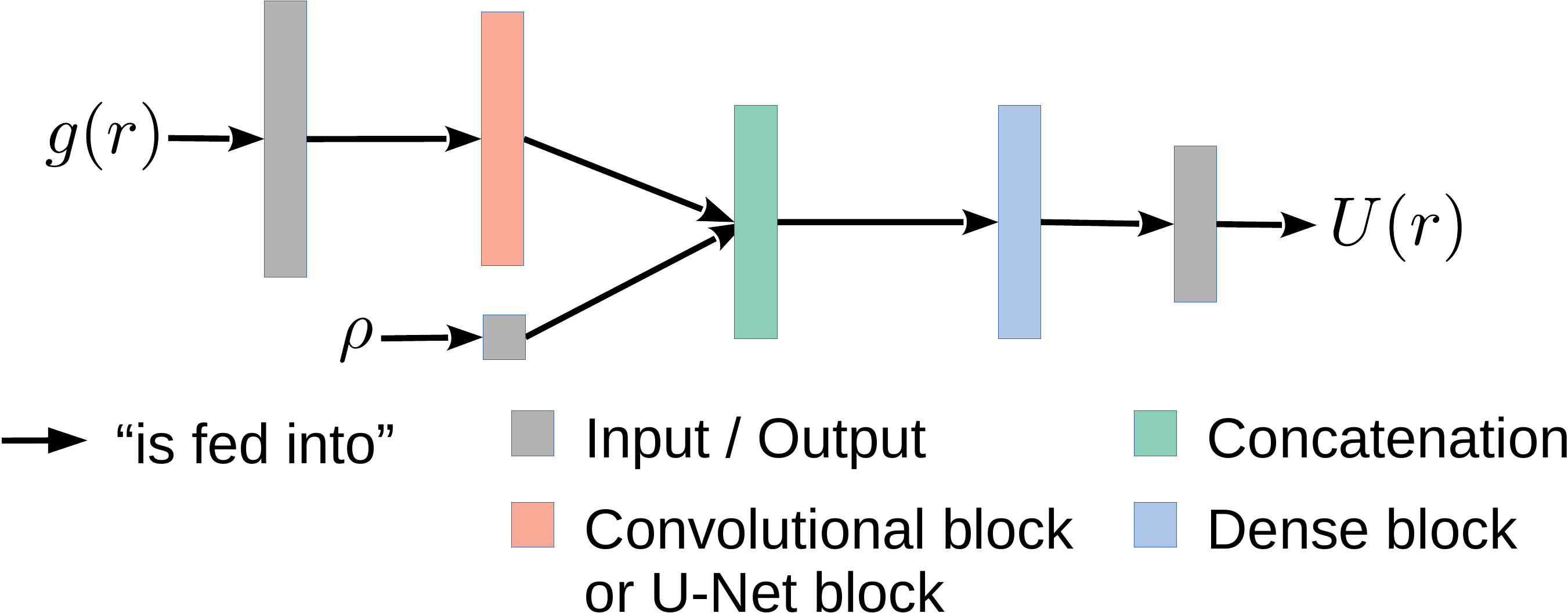}
	\caption{Schematic representation of the NNs used for predicting (effective) pair potentials $U(r)$ from a radial pair distribution function $g(r)$.}
	\label{fig:NN_schematic_PotPredictor}
\end{figure}

In this section we will consider NNs (see Fig.~\ref{fig:NN_schematic_PotPredictor}) for predicting (effective) pair potentials $U(r)$ from radial pair distribution functions $g(r)$ at a given, known particle number density $\rho$. We either used $g(r)$ as input, or computed the (effective) pair potential in the low density limit
\begin{align}
	U_0(r) \coloneqq -k_{\rm B}T \ln\left[g(r)\right] ,
	\label{eq:BI}
\end{align}
and used $U_0(r)$ as the input. The reasoning behind this approach is to provide the NNs with a physically informed estimate which is valid for $\rho \to 0$, so that the NNs only need to learn perturbations to this solution at higher densities. Networks using $U_0(r)$ as input are indicated with a ``b'' for Boltzmann inversion. Optionally, the NNs could optimize and output $\exp[-\beta U(r)]$ instead of $U(r)$ (indicated by ``e''), and/or include the force (indicated by ``f'') in the loss function through Eq.~(\ref{eq:distance-msd}).

The loss was always computed between the predicted, $\hat{U}(r)$, and target pair potential, $U(r)$. To reduce the search space and avoid overfitting, we disregarded parts of $U(r)$ that exceeded $U_{\rm cut} = 20k_{\rm B}T$, effectively yielding a sliced loss function. Hence, the predicted $U(r)$ were not optimized in regions where $U(r) > U_{\rm cut}$, allowing the NNs to set arbitrary values for the corresponding ranges.

Methods such as RMC or IBI instead optimize $\hat{g}(r)$ against the reference $g(r)$ by iteratively performing MD simulations with the predicted potential $\hat{U}(r)$ (see ESI for more details). Such an approach is, however, prohibitively time consuming as it would require additional MD simulations at each step of the optimization procedure. Further, minimizing the loss between $\hat{g}(r)$ and $g(r)$ is infeasible from a technical point of view, because the training of the NNs is based on gradient descent, which requires the (unknown) mapping $U(r) \mapsto g(r)$. Thus, the difference between $g(r)$ and $\hat{g}(r)$ is not a viable input for the loss function for training the NNs, but it can be used as a final performance benchmark in selected instances (see Fig.~\ref{fig:gofrPredictedComparison} below).

\renewcommand{\arraystretch}{1.1}
	\begin{table}[htbp]
 	\begin{tabular}{c|c|c|c}
 		\toprule
 		Abbreviation & $U_0(r)$ input & force in loss & $\exp[-\beta \hat{U}(r)]$ output\\ 
 		\cmidrule[0.4pt](r{0.125em}){1-4}
 		x & \xmark & \xmark & \xmark\\
		\myrowcolour
 		b & \cmark & \xmark & \xmark\\
 		e & \xmark & \xmark & \cmark\\		
		\myrowcolour		
		f & \xmark & \cmark & \xmark\\				
		be & \cmark & \xmark & \cmark\\
		\myrowcolour		
		bf & \cmark & \cmark & \xmark\\	
		ef & \xmark & \cmark & \cmark\\
		\myrowcolour
		bef & \cmark & \cmark & \cmark\\	
 		\bottomrule
 	\end{tabular} 
 	\caption{List of abbreviations for the optional parameters of the networks.}
 	\label{tab:abbreviations_PotPredictor}
 \end{table} 
\renewcommand{\arraystretch}{1.0}

\begin{figure}[htbp]
	\includegraphics[width=\figwidth]{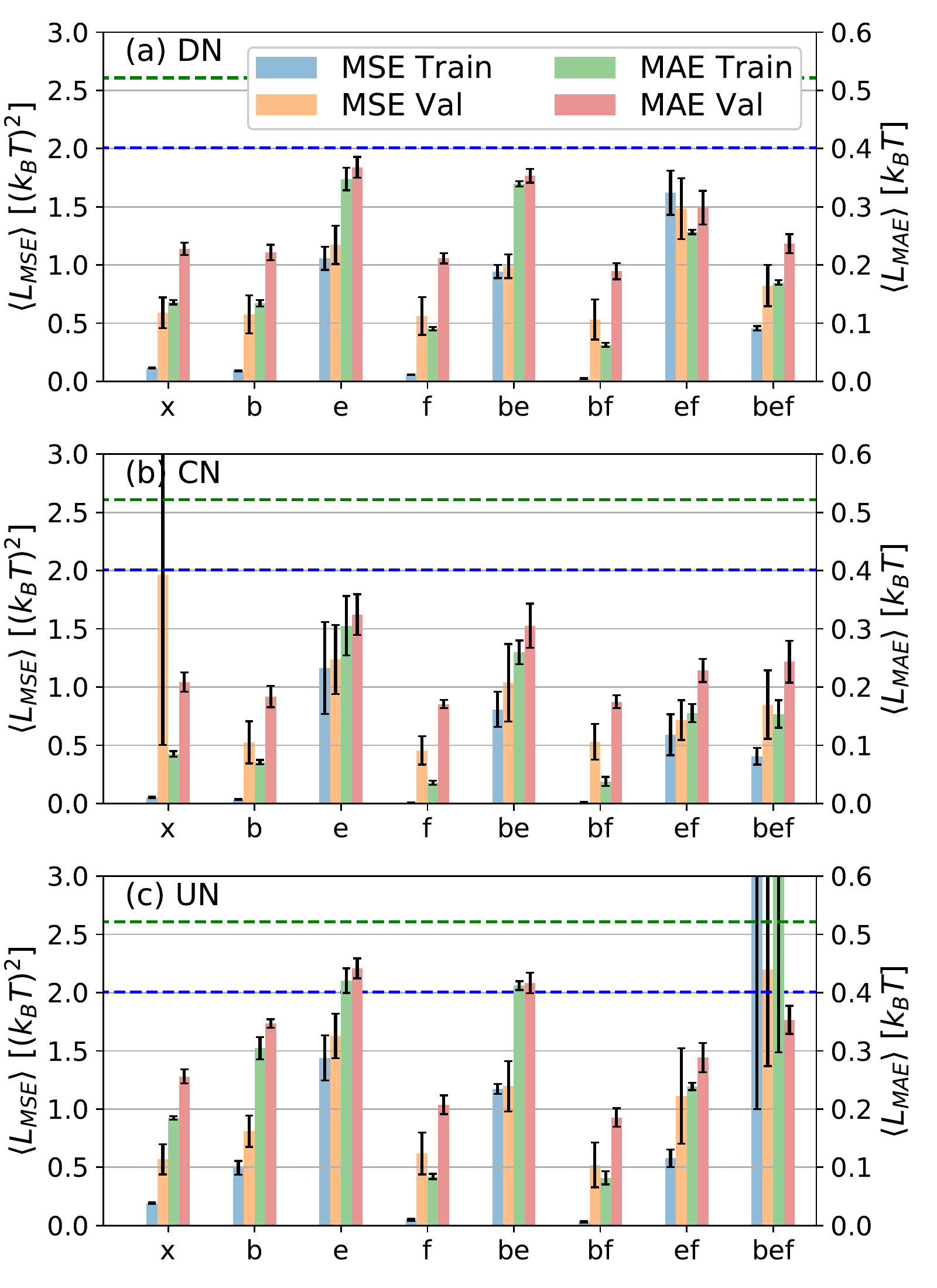}
	\caption{Comparison of $\la L_{\rm MSE} \ra$ (left axis) and $\la L_{\rm MAE} \ra$ (right axis) between the predicted, $\hat{U}(r)$, and target pair potential, $U(r)$, for (a) DNs, (B) CNs, and (c) UNs with different options (see Table~\ref{tab:abbreviations_PotPredictor}). The dashed lines indicate $\la L_{\rm MSE} \ra$ (blue) and $\la L_{\rm MAE} \ra$ (green) between $U(r)$ and $U_0(r)$.}
	\label{fig:potPerformance}
\end{figure}

Again, we tested the DN, CN and UN architectures, and systematically analyzed the influence of providing additional information and different representations of the input and output data. Figure~\ref{fig:potPerformance} shows $\la L_{\rm MSE} \ra$ and $\la L_{\rm MAE} \ra$ between the predicted and the target potentials for all investigated NNs and parameter combinations listed in Table~\ref{tab:abbreviations_PotPredictor}. The errorbars were obtained through a 4-fold cross validation. The dotted lines correspond to the MSE (blue) and MAE (green) between $U(r)$ and $U_0(r)$ [note that we employed here the same clipping as for $\hat{U}(r)$].

\begin{figure}[htbp]
	\includegraphics[width=\figwidth]{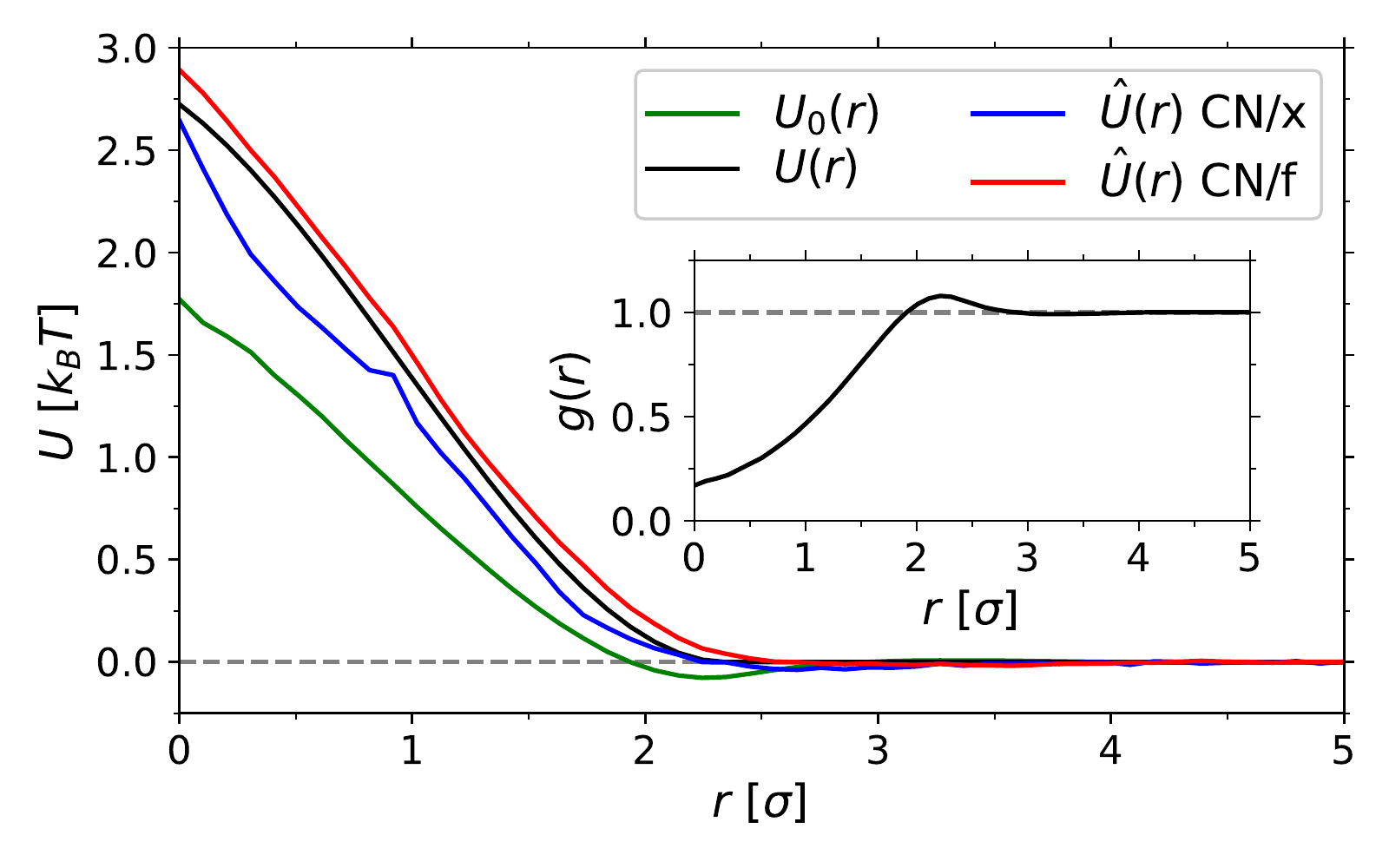}
	\caption{Selected target pair potential $U(r)$ from the validation set and predictions $\hat{U}(r)$ from CN/x and CN/f. The estimated potential $U_0(r)$ is also shown for comparison. The inset shows the corresponding radial distribution function $g(r)$ at the employed density $\rho=0.125\,\sigma^{-3}$.}
	\label{fig:forceComparison}
\end{figure}

As can be seen from Fig.~\ref{fig:potPerformance}, almost all investigated NNs provide more accurate predictions compared to $U_0(r)$. The predicted potentials $\hat{U}(r)$ from the NNs optimizing $\exp[-\beta \hat{U}(r)]$ have consistently larger $\la L_{\rm MSE} \ra$ and $\la L_{\rm MAE} \ra$ compared to the NNs which directly output $\hat{U}(r)$. This discrepancy likely originates from the non-linearity of the transformation, which effectively modifies the importance of different regions of the potential. Figure~\ref{fig:potPerformance} also shows that including the force in the loss function during the training stage improves the accuracy of the NNs (see Sec.~\ref{subsec:NN}). Further, the predicted potentials were significantly smoother when the force was included, as shown for one selected example in Fig.~\ref{fig:forceComparison}. Finally, we find that inputting $U_0(r)$ instead of $g(r)$ did not improve the prediction significantly. Comparing the average losses $\la L_{\rm MSE} \ra$ and $\la L_{\rm MAE} \ra$ of the different network architectures for a given parameter set, we see that the CNs were most accurate in most cases for both the training and validation sets. In particular, the CN/f network had the smallest average losses and also small variations between the different folds. For final benchmarking, we therefore focuses on the CN/f network, which was trained again with all data except those in the test set.

\begin{figure}[htbp]
	\includegraphics[width=\figwidth]{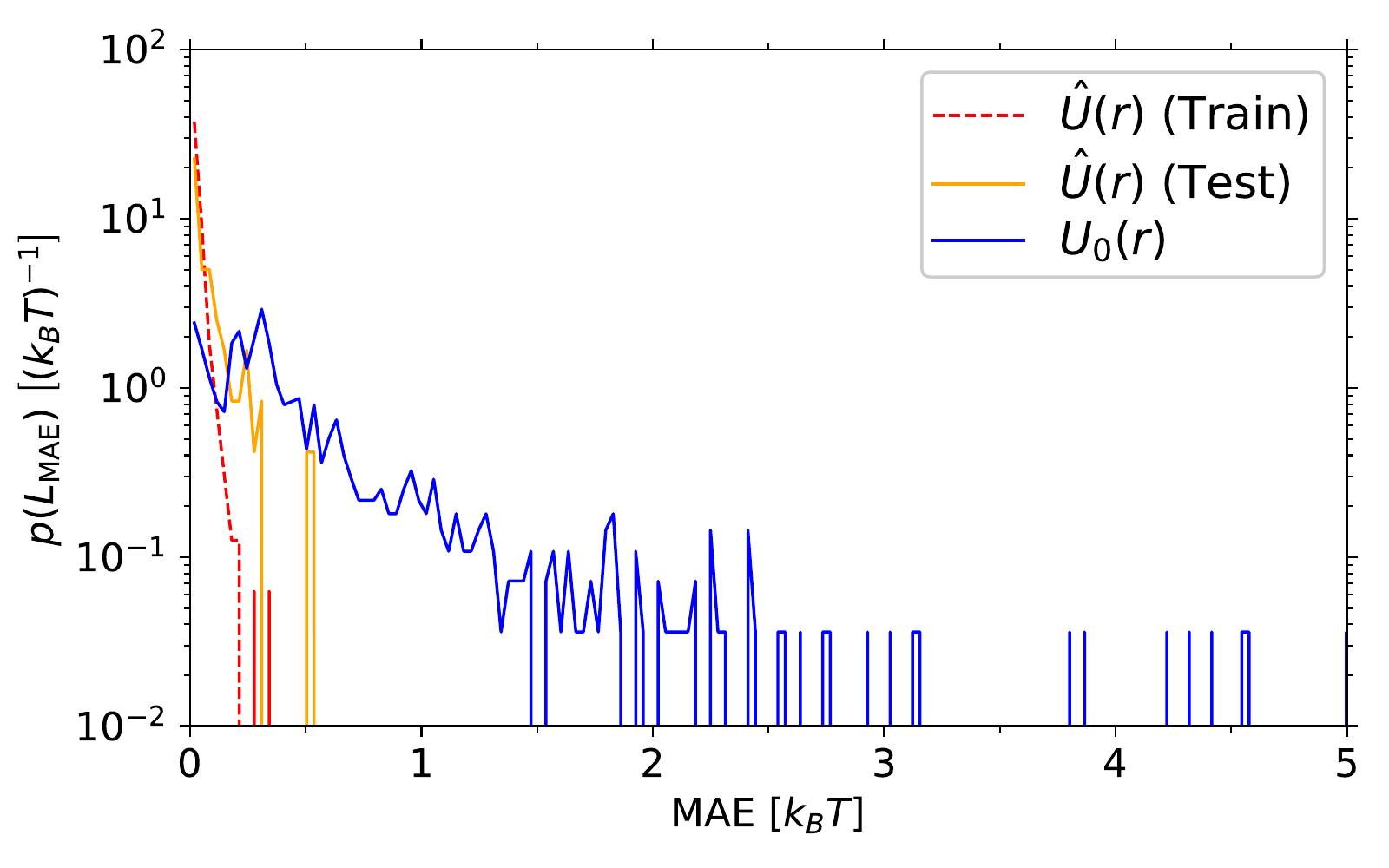}
	\caption{Probability density function of $L_{\rm MAE}$ between target $U(r)$ and predicted $\hat{U}(r)$ using CN/f, and between $U(r)$ and $U_0(r)$.}
	\label{fig:benchmarkPerformance}
\end{figure}

To evaluate the accuracy of this NN in more detail, we analyzed the probability density function of $L_{\rm MAE}$ for the training and test set, $p(L_{\rm MAE})$, and compared it with $p(L_{\rm MAE})$ obtained from using $U_0(r)$ (see Fig.~\ref{fig:benchmarkPerformance}). In both cases, $p(L_{\rm MAE})$ can be fitted by an exponential decay $\propto \exp-(L_{\rm MAE}/\delta)$, with $\delta$ quantifying the width of the distribution. For the CN/f network, $p(L_{\rm MAE})$ had a rather narrow distribution with $\delta = 0.02\,k_{\rm B}T$ and $\delta = 0.03\,k_{\rm B}T$ for the training and test set, respectively, whereas $p(L_{\rm MAE})$ was considerably wider for $U_0(r)$ with $\delta = 0.5\,k_{\rm B}T$ (the asymptotic standard error of the fits was about $10\,\%$ in all cases). It should also be noted that the estimated $U_0(r)$ had several outliers with very large deviations up to $L_{\rm MAE} \approx 8\,k_{\rm B}T$.

\begin{figure}[htbp]
	\includegraphics[width=\figwidth]{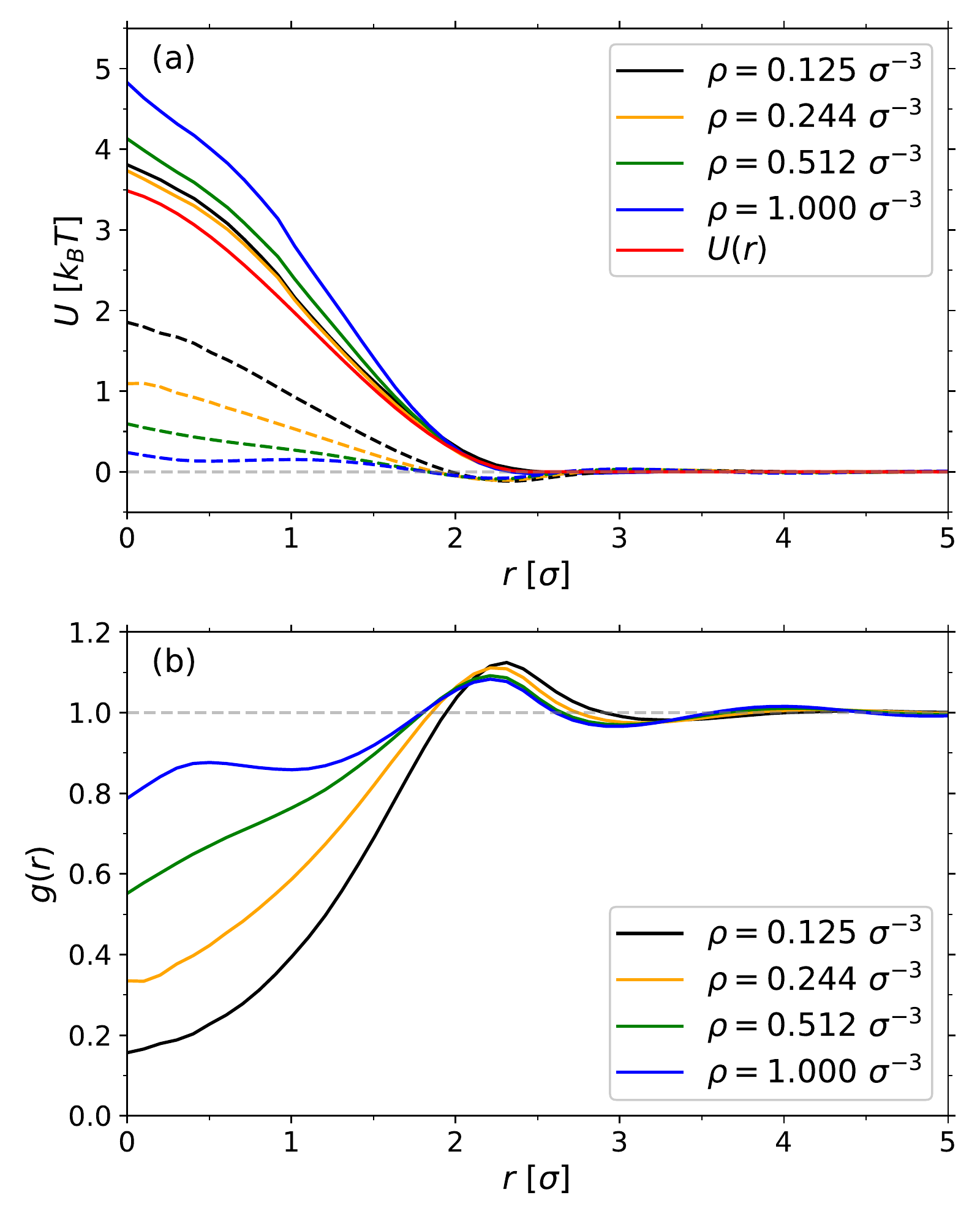}
	\caption{(a) Target potential $U(r)$ and predicted potentials $\hat{U}(r)$ from CN/f at four different particle number densities $\rho$, as indicated. The dashed lines show $U_0(r)$ at the same densities. (b) Radial pair distribution functions $g(r)$ from simulations with $U(r)$ at various $\rho$, as indicated.}
	\label{fig:PotPredictNetComparison}
\end{figure}

At this point, we also want to discuss the transferability of the predicted potentials $\hat{U}(r)$. Given that $g(r)$ can vary significantly for the same $U(r)$ at different particle number densities $\rho$, it is important to check whether $\hat{U}(r)$ remains independent of $\rho$. Figure~\ref{fig:PotPredictNetComparison} shows a selected target potential $U(r)$ from the test set and the resulting $g(r)$ at four different $\rho$, highlighting the increasing deviations from ideal-like behavior [$g(r) = 1$] with increasing density. We have also included in Fig.~\ref{fig:PotPredictNetComparison}(a) the predictions of $\hat{U}(r)$ from the CN/f network as well as $U_0(r)$ at different $\rho$. Indeed, the predictions $\hat{U}(r)$ are rather close to $U(r)$, and show only minor variations with $\rho$ (as should be the case). In contrast, the estimate $U_0(r)$ provides a passable approximation of $\hat{U}(r)$ only at the lowest density $\rho=0.125\,\sigma^{-3}$, and the agreement with $U(r)$ gets significantly worse with increasing $\rho$. 

To check whether the predicted $\hat{U}(r)$ lead to similar pair distribution functions $\hat{g}(r)$ as the original $g(r)$, we ran MD simulations with the predicted potential $\hat{U}(r)$ for selected cases from the training and test set. To test the performance of simple Boltzmann inversion, we also ran MD simulations with $U_0(r)$ and calculated the corresponding $g_0(r)$. Figure \ref{fig:gofrPredictedComparison} shows the comparison between $g(r)$, $\hat{g}(r)$ and $g_0(r)$, where we have also included the employed pair potentials $U(r)$, $\hat{U}(r)$, and $U_0(r)$ as insets of the corresponding plots. In all considered cases, $\hat{g}(r)$ is much closer to $g(r)$ than $g_0(r)$, especially for distances $r \gtrsim \sigma$. The systems shown in Fig.~(\ref{fig:gofrPredictedComparison})(a,b) have been run with pair potentials from the training set at densities $\rho=0.244\,\sigma^{-3}$ and $0.512\,\sigma^{-3}$, respectively. These cases are rather interesting, because $U(r)$ and $\hat{U}(r)$ look almost identical, but the resulting $g(r)$ and $\hat{g}(r)$ show some deviations for $r \lesssim \sigma$. The estimated potentials $U_0(r)$ differ drastically from $U(r)$, but the resulting $g_0(r)$ for $\rho=0.244\,\sigma^{-3}$ is relatively close to the target one. At higher density $\rho=0.512\,\sigma^{-3}$, however, the simulations performed with $U_0(r)$ lead to a distinct clustering of particles, while the simulations with $U(r)$ and $\hat{U}(r)$ lead to an almost flat, gas-like $g(r)$ with weak oscillations. Thus, in this case, $U_0(r)$ would not even be a good starting point for iterative optimization methods like IBI. Figure~(\ref{fig:gofrPredictedComparison})(c,d) shows results for cases from the test set (same systems as in Fig.~\ref{fig:forceComparison}) at two different densities. At low density $\rho=0.244\,\sigma^{-3}$, the agreement between $g(r)$ and $\hat{g}(r)$ is almost perfect, while $g_0(r)$ replicates the qualitative correct trends. For the denser systems $\rho=0.512\,\sigma^{-3}$, both $\hat{g}(r)$ and $g_0(r)$ exhibit a qualitatively similar deviation from the target radial distribution function, but the predicted $\hat{U}(r)$ is much closer to the target pair potential $U(r)$ compared to $U_0(r)$.

\begin{figure*}[htbp]
	\includegraphics[width=\textwidth]{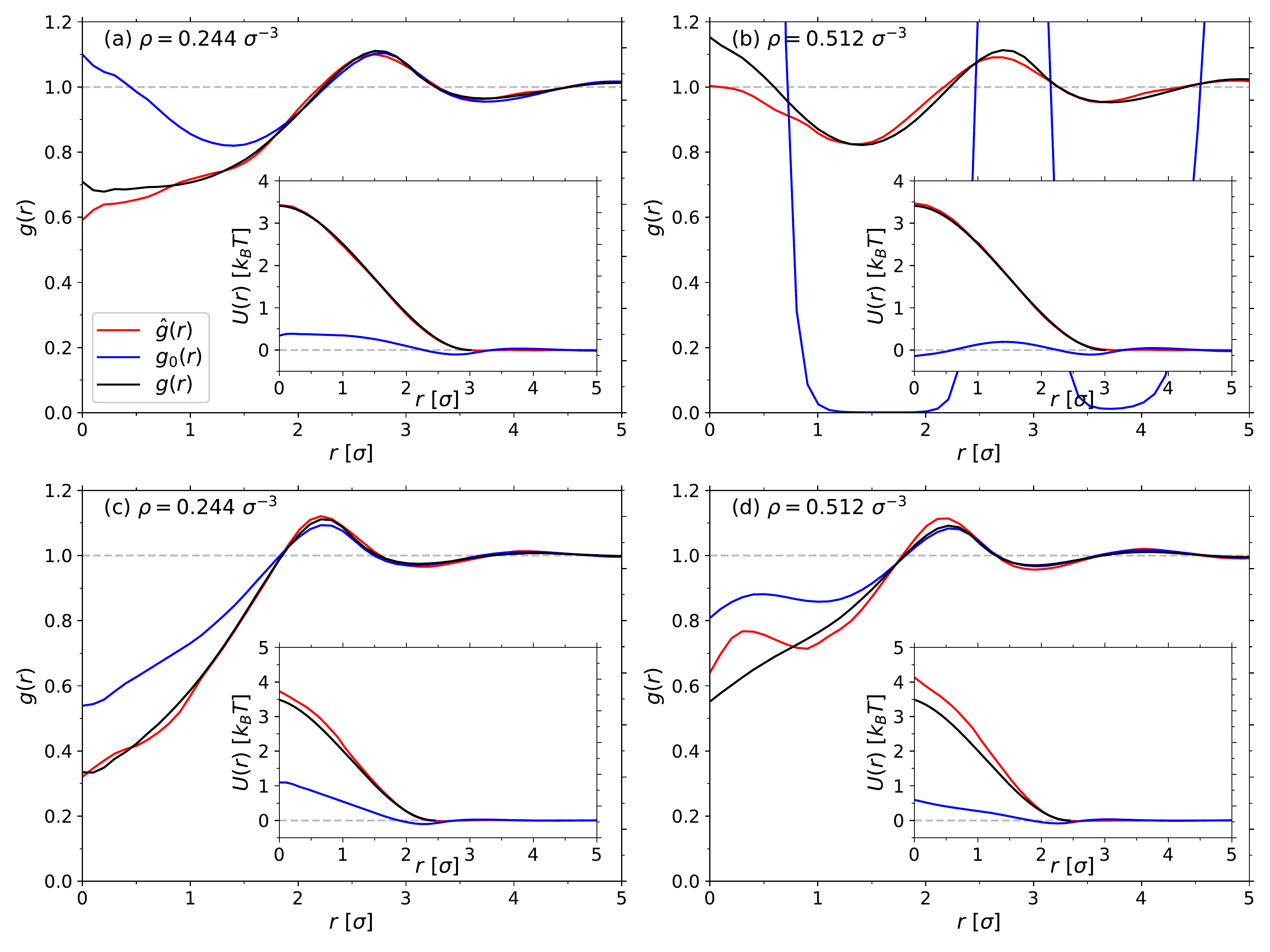}
	\caption{Comparison between radial pair distribution functions $g(r)$, $\hat{g}(r)$, and $g_0(r)$ computed from MD simulations performed with $U(r)$, $\hat{U}(r)$ and $U_0(r)$, respectively. The corresponding potentials have been plotted in the insets. Cases (a,b) are from the training set with (a) $\rho=0.244\,\sigma^{-3}$ and (b) $\rho=0.512\,\sigma^{-3}$, while cases (c,d) are from the test set with (c) $\rho=0.244\,\sigma^{-3}$ and (d) $\rho=0.512\,\sigma^{-3}$. The $L_{\rm MSE}$ between the original $g(r)$ and $\hat{g}(r)$ are (a) $1.4 \times 10^{-4}$ ($5.1 \times 10^{-3}$), (b) $5.0 \times 10^{-4}$ ($5.7 \times 10^{3}$), (c) $2.9 \times 10^{-5}$ ($2.3 \times 10^{-3}$), and (d) $5.0 \times 10^{-4}$ ($2.4 \times 10^{-3}$). The numbers in parentheses indicate $L_{\rm MSE}$ between $g(r)$ and $g_0(r)$.}
	\label{fig:gofrPredictedComparison}
\end{figure*}

\begin{figure}[htbp]
	\includegraphics[width=\figwidth]{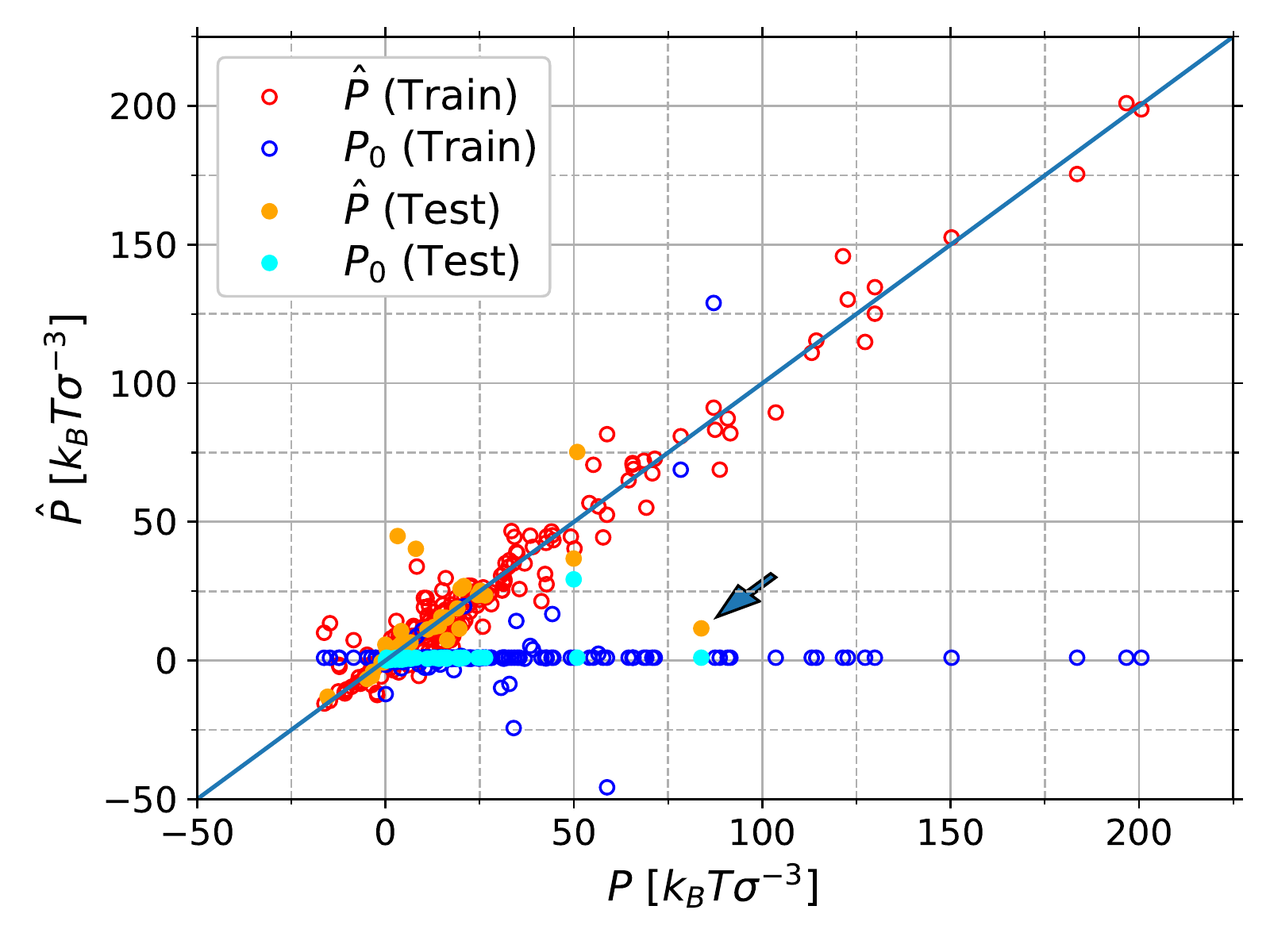}
	\caption{Predicted pressure $\hat{P}$ from UN/c (red and orange symbols) and $P_0$ estimated from Eq.~(\ref{eq:virial_approximation}) (blue and cyan symbols) {\it vs.} the target pressure, $P$, using the potentials $\hat{U}(r)$ and $U_0(r)$, respectively. Open and filled symbols show results from the training and testing set, respectively. The outlier in the test set for the calculation of $\hat{P}$ has been indicated with an arrow.}
	\label{fig:pressureCycle}
\end{figure}

As a final test of our methodology, we used the predicted pair potentials $\hat{U}(r)$ from the CN/f networks as an input of the UN/c networks developed in Sec.~\ref{sec:b2effective} to calculate the pressure $\hat{P}$ from the radial pair distribution function $g(r)$. Figure~\ref{fig:pressureCycle} shows $\hat{P}$ {\it vs.} $P$, demonstrating rather good agreement for the training set ($R^2 \approx 0.971$). The predictions for the test set have a significantly lower accuracy ($R^2 \approx 0.461$), but the overall trends are captured in most cases (excluding the outlier at $P \approx 80\,k_{\rm B}T/\sigma^3$ from the test set leads to $R^2 \approx 0.802$). For comparison, we have also included the pressure $P_0$, which we estimated by computing $U_0(r)$ from $g(r)$. These pressures are almost always much smaller than the target pressure, leading to $R^2 \approx 0.0192$.

\section{Conclusions}
\label{sec:conclusions}
Artificial neural networks (NNs) were developed for predicting the equation of state for a given pair potential $U(r)$ and density $\rho$, and for predicting (effective) pair potentials from structural information based on the radial distribution function, $g(r)$. We investigated how the representation of the input and output data as well as additional information ({\it e.g.}, the forces) influence the prediction accuracy. For both tasks, a key preprocessing step was to limit the input and/or output range by capping the pair potential $U(r)$ so that it could not reach arbitrarily large values. For predicting the pressure $P$, NNs were tested that directly map the input pair potential $U(r)$ to $P$, and NNs that first predict (effective) virial coefficients from $U(r)$ which are then used to compute $P$. The latter strategies resulted in rather accurate predictions, which were much closer to the target pressure ($R^2 \approx 1.00$) compared to the virial expansion derived in the low-density limit. For predicting (effective) pair potentials, the accuracy of the predictions improved significantly when also the derivatives of the potentials were provided in the loss, since more correlations were taken into account. As a result, the predicted potentials became smoother and were significantly closer to the target potentials compared to results obtained through simple Boltzmann inversion.

There is, however, still room for future improvements: the potentials calculated from our NNs reproduce structural data with lower accuracy compared to potentials obtained from iterative methods, such as Iterative Boltzmann Inversion, which are constructed to reproduce $g(r)$. Further, the systems used in this work only covers a subset of all conceivable cases. Nevertheless, our NN approach is still a viable option for obtaining reasonably accurate initial estimates, which can then be optimized further using other methods until the desired accuracy is achieved. Such a combined approach could drastically cut down the computational cost and development time for coarse-graining applications, and it could also be a useful tool for inverse problems in materials discovery in soft matter.

\section*{Supporting Information}
Plots of all potentials and radial distribution functions used for training and testing; Additional information on theoretical background; Schematics of final network architectures

\section*{Acknowledgements}
We thank M. R. Khadilkar for fruitful discussions. This work was funded by the German Research Foundation (DFG) through project number 233630050 - TRR 146. AN further acknowledges financial support provided by the DFG through project number NI 1487/2-1 and NI 1487/2-2. Computing time was granted on the supercomputer Mogon at Johannes Gutenberg University Mainz (www.hpc.uni-mainz.de).

\providecommand{\latin}[1]{#1}
\providecommand*\mcitethebibliography{\thebibliography}
\csname @ifundefined\endcsname{endmcitethebibliography}
  {\let\endmcitethebibliography\endthebibliography}{}

\end{document}